\documentstyle[amsfonts,epsfig,color,11pt]{article}
\makeatletter
\@addtoreset{equation}{section}
\def\theequation{\arabic{section}.\arabic{equation}}
\def\section{\@startsection{section}{1}{\z@}{3.5ex plus 1ex minus
   .2ex}{2.3ex plus .2ex}{\large\bf}}
   \def\thesection{\arabic{section}}

\def\appendix{\setcounter{section}{0}
        \def\thesection{Appendix\ \Alph{section}}
        \def\theequation{\Alph{section}.\arabic{equation}}}

\renewcommand{\thefootnote}{\fnsymbol{footnote}}
\begin{document}
\topmargin 0pt
\oddsidemargin -3.5mm
\headheight 0pt
\topskip 0mm
\addtolength{\baselineskip}{0.68\baselineskip}

\begin{flushright}
\end{flushright}
\vspace{0.5cm}
\begin{center}
    {\large \bf Determining the Minimal Length Scale of the
    Generalized Uncertainty Principle from the Entropy-Area
    Relationship } 
\end{center}
\vspace{0.5cm}
\begin{center}
{Wontae Kim$^{a,c}$\footnote{email:wtkim@sogang.ac.kr} and John
  J. Oh$^{b,d}$\footnote{email:john5@yonsei.ac.kr}}\\[10mm]  

{\small \it {${}^{a}$ Department of Physics, Sogang University, Seoul, 121-742,
Korea\\[0pt] 
${}^{b}$ Department of Physics, Yonsei
    University, Seoul, 120-749, Korea\\[0pt]        

${}^{c}$ Center for Quantum Space-time, Sogang University, Seoul, 121-742,
Korea\\[0pt] 
${}^{d}$ Asia Pacific Center for Theoretical Physics, POSTECH, Pohang, 790-784, Korea\\[0pt]
}}
\vspace{0.3cm}
\end{center}
\begin{center}
    {\small \bf Abstract}
\end{center}
{\small
  We derive the formula of the black hole entropy with a
  minimal length of the Planck size by counting quantum modes of
  scalar fields in the vicinity of the black hole horizon, taking into
  account the generalized uncertainty principle (GUP). This formula is
  applied to some intriguing examples of black holes - the
  Schwarzschild black hole, the Reissner-Nordstrom black hole, and the
  magnetically charged dilatonic black hole.  As a result, it is shown
  that the GUP parameter can be determined by imposing
  the black hole entropy-area relationship, which has a Planck length scale and
  a universal form within the near-horizon expansion.}
 \vspace{1cm}
\begin{flushleft}
{\small PACS numbers: 04.70.Dy, 04.70.-s, 04.62.+v }\\
\end{flushleft}

\newpage
\renewcommand{\thefootnote}{\arabic{footnote}}
\setcounter{footnote}{0}
\section{Introduction}\label{sec:intro}
There has been much interest in the end state
of a small black hole after Hawking evaporation. In the context of
thermodynamics based on the Bekenstein entropy
\cite{Bekenstein} and the Hawking temperature \cite{Hawking:1974sw}, the
black hole emits a radiation and it becomes smaller and hotter, which disappears when the evaporation ends. Then, the black
hole will evaporate completely, leaving behind thermal radiation
described by quantum-mechanical mixed states. Therefore, the
information will be
completely lost and the unitarity postulate of quantum theory may be
broken, which is well-known as {\it the black hole information loss
  paradox}. 
Even though the black hole is charged with an electric and/or magnetic field, the
situation is similar to the uncharged case. There exists a lower bound
of the black hole mass called the extremal limit where the mass and
the charge are in balance. As the small black hole radiates, it looses its
mass and finally approaches the limit at which the black hole radiates no more.

However, this scenario is mainly based on the semi-classical analysis
\cite{Hawking:1976ra}, assuming the 
classical background metric and disregarding
the radiating energy compared to the rest energy of the black hole.
Provided the black hole reaches the Planck size as it radiates, the
emitted radiation energy is not any more negligible compared to the
size of the black
hole.  
Thus, when the size of the black hole is comparable to the Compton wavelength
of the emitted radiation, the quantum fluctuation near the black hole
affects the position of the black hole horizon, which leads
to the breakdown of the semi-classical assumptions. Hence, we should
include the back-reaction effect for the full analysis, which means
that the complete quantum
gravity is required in order to give a definite answer on the fate of
the black hole. 

Recently, it has been proposed that there might be a minimal length with
the Planck scale, modifying usual commutation relations of the Heisenberg's uncertainty principle to the generalized uncertainty principle (GUP). The
tensorial forms of the commutation relation \cite{kempf,gup,nf,kmm,cmot} are 
\begin{eqnarray}
&&[x_{i},p_{j}] = i\hbar (1 + \lambda p^2)\delta_{ij}, \label{eq:gup1}\\
&&[x_{i},x_{j}] = 2i\hbar \lambda (p_{i}x_{j}-p_{j}x_{i}),~~[p_{i},p_{j}] = 0,\label{eq:gup2}
\end{eqnarray}
which leads to the minimal length uncertainty,
\begin{equation}
  \label{eq:gup}
  \Delta x \Delta p \ge \hbar + \frac{\lambda}{\hbar} (\Delta p)^2.
\end{equation}
It has been shown that the commutation relation implying the
minimal length is not uniquely determined \cite{kempf}. Indeed, its
conceptual origin comes from the string theory \cite{gup}, which
resembles noncommutative geometry  \cite{nf,kmm}. In the context of
string theory, the GUP provides the improved 
uncertainty relation (\ref{eq:gup}) and the GUP parameter
$\lambda$ is determined as the fundamental constant associated with the
string tension, $\lambda\simeq \alpha'\sim (10^{-32}{\rm cm})^2$,
implying the existence of the minimal length with a Planck scale,
$\Delta x \ge 2\sqrt{\alpha'}\sim 10^{-32} \rm cm$.

From these reasons, the GUP has drawn much attention in diverse
aspects - the modification of dispersion relations
\cite{kmm,cmot,garay}, the black hole entropy without 
brick walls \cite{li,lius,lhz,sl,kkp,yhk,mkp}, the black hole remnants (BHR) as a
possible resolution of {\it the information loss paradox} \cite{remn}, and the
primordial black hole remnants as a candidate of the cold dark matter
(CDM) \cite{cdm}.  

On the other hand, it has been shown that the GUP relation
(\ref{eq:gup}) can be derived from the model-independent ways from the
quantum theory of gravitation 
\cite{maggiore}, where the GUP parameter has not been specified. However, it can be determined through certain specific models and measurements
such as a string theory or the full theory of quantum gravity.  

One may think the GUP parameter should be determined by some physical
laws or principles because the relation is derived from
the basic assumptions of quantum theory of gravitation. Motivated by this, we would like to
compute the entropy of scalar fields in the black hole background in
the presence of the minimal length, and we show that the black hole
entropy-area relationship can fix the scale of the minimal length of the GUP
in Ref. \cite{maggiore}. More precisely, the generic metric ansatz
with the spherical symmetry is taken into account and the entropy can be written
in the form of the polynomial of the minimal length
parameter in the near-horizon limit. As a result, the most dominant
term describes the entropy that is proportional to the area of the
event horizon while the subleading terms are quite negligible in the
regime of the large black hole.
The GUP parameter determined by the area law is universal up to the
second order expansion of the near-horizon limit in some specific
models, which is the order of the Planck
scale.

In section \ref{sec:entropy}, we shall derive the generic formula of the
entropy of scalar fields in the background of the spherical symmetric
black hole metric assuming semi-classical approximations and keeping
the second order expansion of the near-horizon parameter. In section
\ref{sec:examples}, the charged  
dilatonic black hole solutions with an arbitrary coupling between the dilaton and the U(1) gauge 
field strength are taken into account. For specific values of the
coupling, the solution describes the  
Schwarzschild (SS), the Reissner-Nordstrom (RN), and the magnetically charged dilatonic black 
holes. In section \ref{sec:ent}, for these
specific cases of the coupling, we show that the black hole entropy-area relationship can
determine the GUP parameter as a Planck scale. Finally, some discussions are included in
section \ref{sec:discussion}.

\section{Derivation of Entropy with the Minimal Length}\label{sec:entropy}

Let us consider the Klein-Gordon equation for a massive scalar field
in the background of the black hole $(\Box-\mu^2)\Phi=0$, where $
\Box = \nabla_{\mu}\nabla^{\mu}$ and $\mu$ is the mass of the scalar field.  
Using an ansatz $\Phi = \Psi(r,\theta,\varphi)e^{-i\omega t}$,
then the field equation becomes 
\begin{equation}
\Psi''+
\left(\frac{f'}{f}+2\frac{R'}{R}\right)\Psi'+\frac{1}{f}\left[\frac{\omega^2}{f}-\mu^2
  + \frac{1}{R^2}(\partial_{\theta}^2 + \cot\theta\partial_{\theta} +
  \csc^2\theta \partial_{\varphi}^2)\right]\Psi = 0, 
\end{equation}
where the prime denotes $d/dr$, and $f$ and $R^2$ are the metric functions
of the spherically symmetric metric,
\begin{equation}
(ds)^2 = - f(r)dt^2 + \frac{dr^2}{f(r)} + R^2(r)(d\theta^2+\sin^2\theta d \varphi^2).
\end{equation}

Assuming the Wenzel-Kramers-Brillouin (WKB) approximation
\cite{thooft} with $\Psi \simeq e^{iS(r,\theta,\varphi)}$, it is found to be 
$p_r^2 = \frac{1}{f}\left[\frac{\omega^2}{f}-\mu^2 -
  \frac{p_\theta^2}{R^2} - \frac{p_\varphi^2}{R^2\sin^2\theta}\right], $
where $p_r = \partial S/\partial r$, $p_\theta=\partial
S/\partial\theta$, and $p_\varphi=\partial S/\partial\varphi$. We have
the squared module of momentum given by $p^2 = g^{rr}p_r^2 +
g^{\theta\theta}p_\theta^2+g^{\varphi\varphi}p_{\varphi}^2 =
\omega^2/f - \mu^2$ and the volume in the momentum phase space is
\begin{equation}
V(r,\theta) = \int dp_r dp_\theta dp_\varphi =
\frac{4}{3}\pi\frac{R^2(r)}{\sqrt{f}}\sin\theta
\left(\frac{\omega^2}{f}-\mu^2\right)^{3/2} 
\end{equation}
with $\omega\ge\mu\sqrt{f}$. 
The number of quantum states are given by the weighted phase space volume measure \cite{cmot},
\begin{eqnarray}
n(\omega) &=& \frac{1}{(2\pi)^3} \int dr d\theta d\varphi dp_r dp_\theta dp_\varphi \frac{1}{(1+\lambda p^2)^3} \nonumber \\
  &=& \frac{1}{(2\pi)^3}
\int drd\theta d\varphi \frac{V(r,\theta)}{(1+\lambda p^2)^3}\nonumber\\ 
&=& \frac{2}{3\pi} \int dr \frac{R^2(r)
   \left(\frac{\omega^2}{f}-\mu^2\right)^{3/2}}{\sqrt{f}
\left[1+\lambda\left(\frac{\omega^2}{f}-\mu^2\right)^3\right]}. \label{weight}   
\end{eqnarray} 
One might think  
that arbitrary integration measures can be chosen without deforming
the commutation relations Eqs. (\ref{eq:gup1}) and (\ref{eq:gup2}). 
Specifically, the measure in
Eq. (\ref{weight})
may be absorbed in a suitable rescaling of the fields and in a
suitable redefinition of the operators that act on the fields.  
However, this is not the case since this measure is nontrivial in that  
the measure of the phase space
corresponding to the number of density of states should be
consistently derived as long as we follow the Liouville theorem as indicated in
Ref. \cite{cmot}, where this weighted volume element should be
invariant under the infinitesimal time translations.
Actually, the time evolution is subjected to the GUP through modified
Hamiltonian equations of motion. Of course, taking $\lambda =0$,
then the original density of states is naturally recovered.
The improved measure seems to be plausible in the sense that
the minimal length plays a role of the ultraviolet cut-off 
which is naturally introduced in the
denominator in Eq. (\ref{weight}). 
On the other hand, the construction of the Hilbert space representation with the GUP has
been already done in Ref. \cite{kmm}, 
although it is no longer unique. Indeed, the representation describing
the minimal length uncertainty 
is for the momentum space. Taking into account it in a position
space, the position 
eigenstates are in general no longer orthogonal unlike the momentum
eigenstates. 
Of course, there might be diagonalizable but have no physical
eigenstates since they are described on lattices in the position space. 
In this sense it looks difficult to construct the representation 
in the position space rather than that in the momentum space.
However, for given commutation relations, the weighted volume factor
of the phase space is uniquely determined and is independent of the
choice of representation. 
From these reasons, the invariant phase
space volume gives some corrections 
to the physical quantities through the minimal length
uncertainty principle.

Now, taking into account a thin-layer around the event horizon of the black hole between $r_{+}$ and $r_{+}+\epsilon$, where $\epsilon$ is an infinitesimal distance from the horizon, then the metric functions around this layer are expanded as
\begin{equation}
\label{eq:seriesexp}
f(r)\simeq \kappa (r-r_+) +f_2(r-r_{+})^2,~~R^2(r) \simeq r_0 + r_1
(r-r_+) + r_2 (r-r_+)^2,
\end{equation}
where $\kappa$ is a surface gravity with $\kappa=f'(r_{+})$ and the other
coefficients are $f_{2} = f''(r_{+})/2,~r_0 = R^{2}(r_{+}),~r_{1} =
[R^2(r_+)]'$, and $r_{2}=[R^2(r_+)]''/2$
by keeping the second order of the expansion. Now we want to identify the proper length between the layer with the GUP minimal length. Then the GUP parameter associated with the minimal length $x_{min}=2\sqrt{\lambda}$ is
determined by 
\begin{equation}
\label{eq:pml}
2\sqrt{\lambda} =\int_{x_{+}}^{x_{+}+\epsilon} 
\frac{d\hat{x}}{\sqrt{f(\hat{x})}} \simeq \frac{\sqrt{\epsilon}}{\sqrt{\kappa}}\left(2 - \frac{f_2
    \epsilon}{3\kappa}\right),
\end{equation}
which can be expressed in the alternate form of $\epsilon \simeq \kappa \lambda +
{\mathcal O}(\lambda^2)$. From
the free energy defined by 
\begin{equation}
  \label{eq:freenergy}
F=-\int_{0}^{\infty} d\omega\frac{n(\omega)}{e^{\beta\omega}-1},
\end{equation}
where $\beta$ is the inverse
Hawking temperature, the entropy, ${\mathcal S}_{BH} = \beta^2{\partial F}/{\partial \beta}$, is straightforwardly calculated as 
\begin{eqnarray}
{\mathcal S}_{BH} &=& \frac{\beta^3}{12\pi\lambda^3}\int_{0}^{\infty}dx
\frac{x^4}{\sinh^2x}\int_{r_{+}}^{r_{+}+\epsilon} dr
\frac{R^2(r)f}{(x^2+Bf)^3}\nonumber \\&\equiv&
\frac{\beta^3}{12\pi\lambda^3}\int_{0}^{\infty}dx
\frac{x^4}{\sinh^2x}{\mathcal I}(x), \label{eq:entrop} 
\end{eqnarray}
where $B=\beta^2/4\lambda$ and $x\equiv
\beta\omega/2$ by setting
$\mu=0$ (massless scalar field) for simplicity. 
Since we have
  $R^2 f = r_0\kappa (r-r_+) + (r_1\kappa+r_0f_2)(r-r_+)^2 +
  (r_2\kappa+r_1f_2)(r-r_+)^3 +{\mathcal O}(r-r_+)^4$,
the radial integration, ${\mathcal I}(x)$ in Eq. (\ref{eq:entrop}) is expressed by three parts near the horizon and expanded in terms of $\epsilon$ by keeping the $\epsilon^4$-order terms, 
\begin{equation}
{\mathcal I}(x) = \int_{0}^{\epsilon} d\hat{\epsilon} \frac{r_0\kappa
  \hat{\epsilon} + (r_1\kappa+r_0f_2)
  \hat{\epsilon}^2 + (r_2\kappa + r_1 f_2)  \hat{\epsilon}^3}{(x^2 + B\kappa\hat{\epsilon}+Bf_2
  \hat{\epsilon}^2)^3} \simeq \frac{a}{x^6} - \frac{b}{x^8} + \frac{c}{x^{10}},
\end{equation}
where
\begin{eqnarray}
  a &=& \frac{1}{2}r_0\kappa \epsilon^2 +
  \frac{1}{3}(r_1\kappa+r_0f_2)\epsilon^3 +
  \frac{1}{4}(r_2\kappa+r_1f_2)\epsilon^4,\label{eq:as1}\\
  b &=& \frac{\beta^2\kappa\epsilon^3}{16\lambda}\left[4r_0\kappa +
  3(2r_0f_2 + r_1\kappa)\epsilon\right],~~c =
  \frac{3r_0\kappa^3\beta^4 \epsilon^4}{32\lambda^2}.\label{eq:as2}
\end{eqnarray}
Therefore, the black hole entropy becomes 
\begin{equation}
  \label{eq:bentropy}
  {\mathcal S}_{BH} = \frac{\beta^3}{12\pi \lambda^3}
  \left[ a \int_{0}^{\infty} \frac{dx}{x^2 \sinh^2 x} - b
  \int_{0}^{\infty} \frac{dx}{x^4 \sinh^2 x} + c
  \int_{0}^{\infty} \frac{dx}{x^6 \sinh^2 x}\right].
\end{equation}
Since the integrations with respect to $x$ can be regarded as a contour
integration on a complex plane, we use the residue theorem and find 
\begin{equation}
  \label{eq:contour}
  \int_{0}^{\infty} \frac{dx}{x^2\sinh^2 x} = \frac{2}{\pi^2}\zeta(3),~~
  \int_{0}^{\infty} \frac{dx}{x^4\sinh^2 x} = -\frac{4}{\pi^4}\zeta(5),~~
  \int_{0}^{\infty} \frac{dx}{x^6\sinh^2 x} = \frac{6}{\pi^6}\zeta(7),
\end{equation}
where $\zeta(n)$ is a zeta function. Plugging these into Eq. (\ref{eq:bentropy}), then the black hole entropy is
\begin{equation}
  \label{eq:entropy}
  {\mathcal S}_{BH} = \frac{\beta^3}{12\pi^3\lambda^3} \left[ 2a
  \zeta(3) + \frac{4b}{\pi^2}\zeta(5) + \frac{6c}{\pi^4} \zeta(7)\right],
\end{equation}
which is the general formula of the entropy. Since the precise form of the black hole entropy depends on the specific metric, we shall apply this to some concrete black hole models in the following section.

\section{Dilatonic Charged Black Holes with an Arbitrary Coupling}\label{sec:examples}

Now let us consider a four-dimensional low-energy dilaton gravity
action with an arbitrary coupling from string
theory given by \cite{ghs}
\begin{equation}
\label{eq:action}
I=\frac{1}{16\pi G_{N}}\int d^4x \sqrt{-g}\left[{\mathcal{R}}-2(\nabla\phi)^2 - e^{-2\alpha \phi}F^2\right],
\end{equation}
where $G_{N}$ is a Newton's constant, $\phi$ is a dilaton field, $F$ is a Maxwell field strength of
a $U(1)$ subgroup of $E_8\times E_8$ or ${\rm Spin}(32)/Z_2$, and
$\alpha$ is a coupling constant between dilaton and the Maxwell field
strength. The charged dilatonic black hole solution with a spherical
symmetry is given in the form of 
\begin{eqnarray}
f(r) &=& \left(1-
  \frac{r_{+}}{r}\right)\left(1-\frac{r_{-}}{r}\right)^{\frac{1-\alpha^2}{1+\alpha^2}},~~R^2(r) = r^2\left(1-\frac{r_{-}}{r}\right)^{\frac{2\alpha^2}{1+\alpha^2}},\\e^{-2\alpha\phi} &=&
\left(1-\frac{r_{-}}{r}\right)^{\frac{2\alpha^2}{1+\alpha^2}},~~F=Q
\sin\theta d\theta \wedge d\varphi 
\end{eqnarray}
where $r_{+}$ and $r_{-}$ are related to the mass $M$ and the magnetic charge $Q$ of black holes as
$2M=r_{+} + \frac{1-\alpha^2}{1+\alpha^2}r_{-},~~Q^2 = \frac{r_{+}r_{-}}{1+\alpha^2}$, respectively.
It is easy to verify that the action (\ref{eq:action}) has an
electro-magnetic dual symmetry under $\phi \rightarrow
-\phi$ along with the fixed metric by defining the electric field
strength as $\tilde{F}_{\mu\nu} = \frac{1}{2}e^{-2\alpha\phi} \epsilon_{\mu\nu}^{~~~\rho\sigma}F_{\rho\sigma}$.
The solution for $\alpha=0$ describes the Ressiner-Nordstrom (RN)
black hole while the one for $\alpha=1$ corresponds to the magnetically
charged black hole \cite{gm}. Moreover, the case of $\alpha=0$ and
$r_{-}=0$ describes the Schwarzschild black hole. 
Note that the extremal limit $r_{+}=r_{-}$ leads to $M^2=Q^2/(1+\alpha^2)$.
The Hawking temperature for non-extremal black holes is given by $
T_{H} = \beta^{-1}=\frac{1}{2\pi
  r_{+}}\left(\frac{r_{+}-r_{-}}{r_{+}}\right)^{\frac{1-\alpha^2}{1+\alpha^2}}$,
which implies that it always vanishes for the extremal limit unless $\alpha=1$.

\section{Universal Minimal Length Scale from the
  Entropy-Area Relationship}\label{sec:ent}
\subsection{Schwarzschild (SS) and Reissner-Nordstrom (RN) black holes}\label{sec:SSRN}

For the SS black hole ($\alpha=0$ and $r_{-}=0$), the coefficients of
the series expansion in Eq. (\ref{eq:seriesexp}) can be identified with
$\kappa={1}/{r_{+}}$, $f_2=-{1}/{r_{+}^2}$, $r_0=r_{+}^2$, $r_1 =
2r_{+}$, and $r_2=1$ and then Eqs. (\ref{eq:as1}) and (\ref{eq:as2}) become 
\begin{equation}
a = \frac{r_{+}\epsilon^2}{2} +\frac{\epsilon^3}{3} -
\frac{\epsilon^4}{4r_{+}},~~b = \frac{\beta^2\kappa\epsilon^3
  r_{+}}{4\lambda},~~c=\frac{3\beta^4\epsilon^4}{32\lambda^2
  r_{+}}. 
\end{equation}
Plugging these into Eq. (\ref{eq:entropy}), one finds the black hole
entropy in the polynomial form with respect to $\lambda$ as 
\begin{equation}
\label{eq:entropySS}
{\mathcal S}_{BH}^{SS} = \frac{A_{H}}{4G_{N}} + \frac{4}{9}\zeta(3) + {\mathcal O}(\lambda),
\end{equation}
where $A_{H}$ is the area of $S^{2}$-sphere at the event horizon,
$A_{H}=4\pi r_{+}^2$, and the 
GUP parameter is determined to $\lambda = 2G_{N}
(\zeta(3)+4\zeta(5)+9\zeta(7))/3\pi$. Note that the GUP parameter depends on the Newton's
constant of the Planck length scale since $\lambda \sim 3.061061275 \times G_{N}$.
The leading order describes
the area law of the black hole entropy with the $1/\lambda$-order
contribution while the next leading term is the $\lambda^{0}$-th and higher
order terms. However, in the large
black hole limit we considered ($r_{+}>>1$), the subleading terms are
quite negligible compared to the area term.

On the other hand, for the RN black hole, the
coefficients for the expansion of the metric functions can be given by
to $\kappa=(r_{+}-r_{-})/r_{+}$, $f_2 = - (r_{+}-2r_{-})/r_{+}^3$,
$r_0=r_{+}^2$, $r_1 = 2 r_{+}$, and $r_{2} = 1$ and one can easily find
\begin{eqnarray}
&&a = \frac{(r_{+}-r_{-})\epsilon^2}{2} + \frac{\epsilon^3}{3}
-\frac{(r_{+}-3r_{-})\epsilon^4}{4r_{+}^2},\nonumber \\
&&b = \frac{\beta^3\kappa\epsilon^3}{8\lambda}\left(2(r_{+}-r_{-})
  + \frac{3r_{-}\epsilon}{r_{+}}\right),~c =
\frac{3\beta^4\epsilon^4 (r_{+}-r_{-})^3}{32\lambda^2 r_{+}^4}. 
\end{eqnarray}
Therefore, one can show the entropy of the RN black hole up to the zeroth order of $\lambda$,
\begin{equation}
\label{eq:entropyRN}
{\mathcal S}_{BH}^{RN} = \frac{A_{H}}{4G_{N}} + \frac{4}{9}\zeta(3) +
\frac{4r_{-}}{r_{+}}\zeta(5) + {\mathcal O}(\lambda), 
\end{equation}
where the same GUP parameter $\lambda$ as the case of the SS black hole was chosen. The first term
represents the area law of the black hole entropy, which agrees
with the result in Ref. \cite{yhk} while the higher order corrections
are negligible for the large black hole case.

\subsection{Magnetically charged dilatonic black hole}\label{sec:MCDBH}

The magnetically charged dilatonic black hole solution is obtained
when $\alpha=1$, which yields $f(r)=1-r_{+}/r$ and $R^2(r) =
r(r-r_{-})$. For this metric solution, the coefficients of the series expansion
near the horizon are easily obtained as
$\kappa={1}/{r_{+}},~f_2=-{1}/{r_{+}},~r_{0}=r_{+}(r_{+}-r_{-}),~r_{1}=2r_{+}-r_{-},~r_{2}=1$,
and the coefficients (\ref{eq:as1}) and (\ref{eq:as2}) are found to
be 
\begin{eqnarray} 
 && a= \frac{(r_{+}-r_{-})\epsilon^2}{2}+\frac{\epsilon^3}{3}
  -\frac{(r_{+}-r_{-})\epsilon^4}{4r_{+}^2},\nonumber\\
  &&b= \frac{\beta^2\epsilon^3}{16\lambda r_{+}}
  \left(4(r_{+}-r_{-})+\frac{3r_{-}\epsilon}{r_{+}}\right),
  ~~c=\frac{3\beta^4\epsilon^4(r_{+}-r_{-})}{32\lambda^2 r_{+}^2}.
\end{eqnarray}
Hence, the entropy of the charged dilatonic black hole is found by keeping
the $\lambda^{0}$-th order,
\begin{equation}
  \label{eq:entdbh}
  {\mathcal S}_{BH}^{\alpha=1} = \frac{A_{H}}{4G_{N}} +
  \frac{4}{9}\zeta(3) + \frac{2r_{-}}{r_{+}}\zeta(7) + {\mathcal O}(\lambda),
\end{equation}
where $A_{H} = 4\pi r_{+}(r_{+}-r_{-})$ and $\lambda =
2G_{N}(\zeta(3)+4 \zeta(5)+9\zeta(7))/3\pi$. Note that the
first leading term represents the area of $S^{2}$-sphere at the horizon
while the next subleading term is also negligible for the limit of
$r_{+}>>1$. Therefore, the GUP parameter of the Planck scale has an universal form,
as seen in the previous two cases up to the second order
for the near-horizon expansion.

\section{Discussion}\label{sec:discussion}

We have derived the generic formula of the entropy of black holes by integrating quantum modes of scalar fields, taking into account the modified dispersion relation from the
GUP and including the next subleading terms of the near-horizon expansion. This is a general result in that it is independent of the metric
solutions as long as we assume the spherical symmetry. Since we identified the
ultra-violet cut-off $\epsilon$ with the
GUP parameter $\lambda$, one can expand the metric function with
respect to this GUP parameter and the expansion is also valid with respect to the minimal length of the Planck scale.
The generic formula have been applied to the
Schwarzschild black hole, the RN black hole, and the magnetically
charged dilatonic black hole. The leading term of the entropies in
three cases is the $\lambda^{-1}$-order, which clearly describes the area law of the black hole,
whereas the subleading terms are negligible compared to the leading
term since we have used the
semi-classical assumptions that is valid only for the large black hole
case. Especially, the constant contribution to the entropy can be
removed by an appropriate normalization.  

A short glance of three exemplified results reveals that
the scale of the GUP parameter can be determined by the entropy-area
relationship, which has the Planck length scale. Thus the
unpredictable GUP parameter in Ref. \cite{maggiore} can be fixed by the
black hole entropy-area relationship, which is universal up
to the second order expansion of the near-horizon limit.
Therefore, we conclude that the physical scale of the GUP parameter
derived from the quantum theory of gravitation in Ref. \cite{maggiore} can
be predicted by the black hole entropy-area relationship as 
$\lambda = \frac{2}{3\pi} (\zeta(3)+4\zeta(5)+9\zeta(7)) G_{N} \sim G_{N} 
$,
 which has a Planckian scale of the minimal length, $\Delta x = 2\sqrt{\lambda} \sim 10^{-33}~{\rm cm}$.

The minimal length uncertainty has been naturally derived from various 
contexts such as the string theory and the quantum gravity, 
which advocate the fundamental features of the UV/IR correspondence.  
One may expect that there is no invariance under the Lorentz 
transformation at the Planck scale where the minimal length 
uncertainty is dominant. In fact, the Lorentz symmetry breaking 
at the string scale or the Planck scale has been suggested in
Ref. \cite{suss}. Furthermore, a similar possibility appears 
in the formalism of the $\kappa$-deformed Poincare group \cite{lnr} 
since this quantum deformation closely related to the GUP  
also breaks the Lorentz invariance \cite{magg2}. 
The minimal
length is of great interest since 
it exhibits some intriguing feature of the UV/IR relation in a variety
of contexts such as the AdS/CFT correspondence \cite{adscft},
noncommutative field 
theories \cite{ncft}, quantum gravity in asymptotically de Sitter
space \cite{ds}, and so on. 
In spite of these nice non-relativistic arguments, it should be possible to obtain 
the Lorentz covariant formulation of the GUP, which is unfortunately not successful
up to now. For instance, the GUP commutation relations 
(\ref{eq:gup1}) and (\ref{eq:gup2}) are not fully tensorial
forms, which clearly breaks the Lorentz covariance.
So, one may try to single out a preferred frame from a Lorentz covariant formulation. For this purpose, 
let us simply write the GUP relations in a Lorentz covariant fashion,
$[x_{\mu},p_{\nu}] = i\hbar (1+\lambda p_{\alpha}p^{\alpha}) 
\eta_{\mu\nu},~~
[x_{\mu},x_{\nu}] = 2i\hbar \lambda (p_{\mu}x_{\nu} -
x_{\mu}p_{\nu}),~~[p_{\mu},p_{\nu}] = 0$.
Then, the usual GUP commutation relations are recovered 
by dropping term $(p^0)^2$ from $p_{\alpha}p^{\alpha}$. However,
this procedure is not equivalent to the non-relativistic limit
\cite{qt} although
the commutative limit can be well-defined for $\lambda=0$.
It implies that the above Lorentz covariant formulation fails so that
the preferred frame giving the minimal length cannot be found from
this naive formulation. Historically, as the relativistic quantum
mechanics from the old quantum mechanics is not straightforward,
it seems that it is not easy to achieve the relativistic formulation of the
GUP which is the nontrivial extension of the quantum mechanics. 
We hope that this intriguing and important problem 
will be studied elsewhere.

\vspace{0.7cm}
\textbf{Acknowledgments}

{
J. J. Oh would like to thank Robert B. Mann, Mu In Park, Hyeong-Chan
Kim, Hyeonjoon Shin, and Edwin J. Son for useful discussions.  
J. J. Oh is also grateful to Sang Pyo Kim and the headquarter of APCTP for warm hospitality during
the APCTP-TPI Joint Focus Program and Workshop.
We would like to thank Seungjoon Hyun for helpful discussions and comments. 
J. J. Oh was supported by the Brain Korea 21(BK21) project funded by
the Ministry of Education and Human Resources of Korea Government. 
W. Kim was in part supported by the Sogang Research Grant, 20071063 (2007) and 
the Science Research Center Program of the Korea Science and Engineering Foundation through the Center for Quantum Spacetime (CQUeST) of Sogang University with grant number R11-2005-021, and  also in part supported by the Korea Science and Engineering Foundation (KOSEF) grant funded by the Korea government(MOST) (R01-2007-000-20062-0).}
\vspace{0.7cm}

\end{document}